\documentclass[10pt,journal]{IEEEtran}
\usepackage{amsmath,graphicx}
\usepackage{amsfonts}
\usepackage{cite}
\usepackage{multicol}
\usepackage{caption}
\usepackage{booktabs}
\usepackage{subfigure}
\usepackage{url}
\usepackage{setspace}      
\usepackage{multirow}
\usepackage{stfloats}
    \DeclareMathOperator{\sech}{sech}
\title{Detecting Gas Vapor Leaks Using Uncalibrated Sensors}
\author{
    \IEEEauthorblockN{Diaa Badawi\IEEEauthorrefmark{1} \thanks{This work is being supported in part by NSF grants 1739396 (UIC), 1739451 (ASU), 1739390 (UD) and 1739684 (UCSD) .  The authors, Badawi and Cetin, additionally thank Nvidia for an equipment grant.}Tuba Ayhan\IEEEauthorrefmark{2} Sule Ozev\IEEEauthorrefmark{3} Chengmo Yang\IEEEauthorrefmark{4} Alex Orailoglu\IEEEauthorrefmark{5}  A. Enis \c{C}etin\IEEEauthorrefmark{1}}
    \\
    \IEEEauthorblockA{\IEEEauthorrefmark{1}Department of Electrical and Computer Engineering, University of Illinois at Chicago, Illinois}\\
    \IEEEauthorblockA{\IEEEauthorrefmark{2}Electronics and Communication Engineering Department, Istanbul Technical University, Turkey}\\
    \IEEEauthorblockA{\IEEEauthorrefmark{3}School of Electrical, Computer and Energy Engineering, Arizona State University, Tempe, AZ}\\
    \IEEEauthorblockA{\IEEEauthorrefmark{4}Department of Electrical \&  Computer Engineering, University of Delaware, Newark, DE} \\
    \IEEEauthorblockA{\IEEEauthorrefmark{5}Department of Computer Science \& Engineering, University of California, San Diego, La Jolla, CA}\\
    Emails: \IEEEauthorrefmark{1}\{dbadaw2, aecyy\}@uic.edu, \IEEEauthorrefmark{2}ayhant@itu.edu.tr, \IEEEauthorrefmark{3}sule.ozev@asu.edu, \IEEEauthorrefmark{4}chengmo@udel.edu, \IEEEauthorrefmark{5}alex@cs.ucsd.edu}

\begin{document}
%\ninept
%
\maketitle
\begin{abstract}
Chemical and infra-red sensors generate distinct responses under similar conditions because of sensor drift, noise or resolution errors. In this work, we use different time-series data sets  obtained by infra-red and E-nose sensors in order to detect Volatile Organic Compounds (VOCs) and Ammonia vapor leaks. We process time-series sensor signals using deep neural networks (DNN). Three neural network algorithms are utilized for this purpose. Additive neural networks (termed AddNet) are based on a multiplication-devoid operator and consequently exhibit energy-efficiency compared to regular neural networks. The second algorithm uses generative adversarial neural networks so as to expose the classifying neural network to more realistic data points in order to help the classifier network to deliver improved generalization. Finally, we use conventional convolutional neural networks as a baseline method and compare their performance with the two aforementioned deep neural network  algorithms in order to evaluate their effectiveness empirically. 
%In this paper, we describe a Cyber-Physical System (CPS) to detect gas vapor leaks.
%While Volatile Organic Compounds (VOC) and ammonia have a place in our daily lives, their leakage into the environment is harmful to human health. In order to prevent and detect gaseous leaks of harmful VOCs, a cyber-physical system (CPS) comprised of ordinary people or first responders is proposed. This CPS uses small, low-cost sensors coupled to smart phones or
%energy efficient 
%mobile devices with the necessary computation and communication capabilities. 
%The efficacy of such a CPS  hinges on its ability to address technical challenges stemming from the fact that identically produced sensors may produce different results under the same conditions due to sensor drift, noise, or resolution errors. 

%The proposed system makes use of time-varying signals produced by sensors to detect gas leaks. Sensors sample the gas vapor level in a continuous manner and time-varying sensor data is processed using deep neural networks. One of the neural networks (NN) is an energy efficient Additive Neural Network (AddNet) which can be implemented in host devices. The second NN is the discriminator of a GAN and the third a regular convolutional NN. AddNet produces comparable VOC gas leak detection results to regular convolutional networks while reducing area requirements by two thirds. 
\end{abstract}

%\vspace{-0.1cm}
\section{Introduction}
\label{sec:intro}
%\vspace{-0.1cm}
%\renewcommand{\baselinestretch}{1.0}
%\doublespacing
Ammonia and Volatile organic compounds (VOCs) are associated with numerous health problems. Although VOCs and Ammonia are naturally occurring, they can nonetheless cause serious health issues in high concentration. For example, exposure to Ammonia in high concentration causes harm to skin, lungs and eyes. Methane and other VOC compound leaks contribute to global warming. VOC compounds such  as  benzene and toluene are carcinogenic\cite{jones1999indoor,brown1994concentrations,maltoni1989benzene,kang2017indoor}. 

%Volatile organic compounds (VOCs) and ammonia can be harmful to human health. If their indoor concentration exceeds a certain level, they   may trigger asthma and rhinitis.  VOC gases are a major contributor to global warming. Additionally, some VOC compounds such as benzene and toluene are carcinogenic. High concentrations of ammonia pose a health hazard as well as cause harm to the skin, eyes, and lungs\cite{jones1999indoor,brown1994concentrations,maltoni1989benzene,kang2017indoor}. 
%Exposure to 300 ppm or higher is dangerous to life and health. 
%Exposure to ammonia can occur from a deliberate terrorist attack or through an accidental release from farms or industrial and commercial facilities.

In this paper, we consider both an infrared (IR) and a chemical sensor system for early detection and, thus, prevention of dangerous gas leaks\footnote{This work was presented in part at the 2019 IEEE International Conference on Acoustics, Speech and Signal Processing (ICASSP), Brighton, UK, May 2019 \cite{badawi2019detecting}.}. Mobile infrared and chemical sensors can be part of an open air cyber-physical system (CPS)\cite{badawi2019detecting}. We use the time-series data obtained by the sensors in order to detect accidental and/or deliberate gas vapor leaks. The main contribution of this paper centers on the exploitation of the time-series data that sensors produce rather than conventional reliance on a single or a couple of sensor readings for leak detection.
%These processing node can be mobile processors or smart phones. Finally, the human agents will be alerted in case of gas leakage and will take action accordingly.

\begin{figure}
    \centering
    \includegraphics[width=\linewidth]{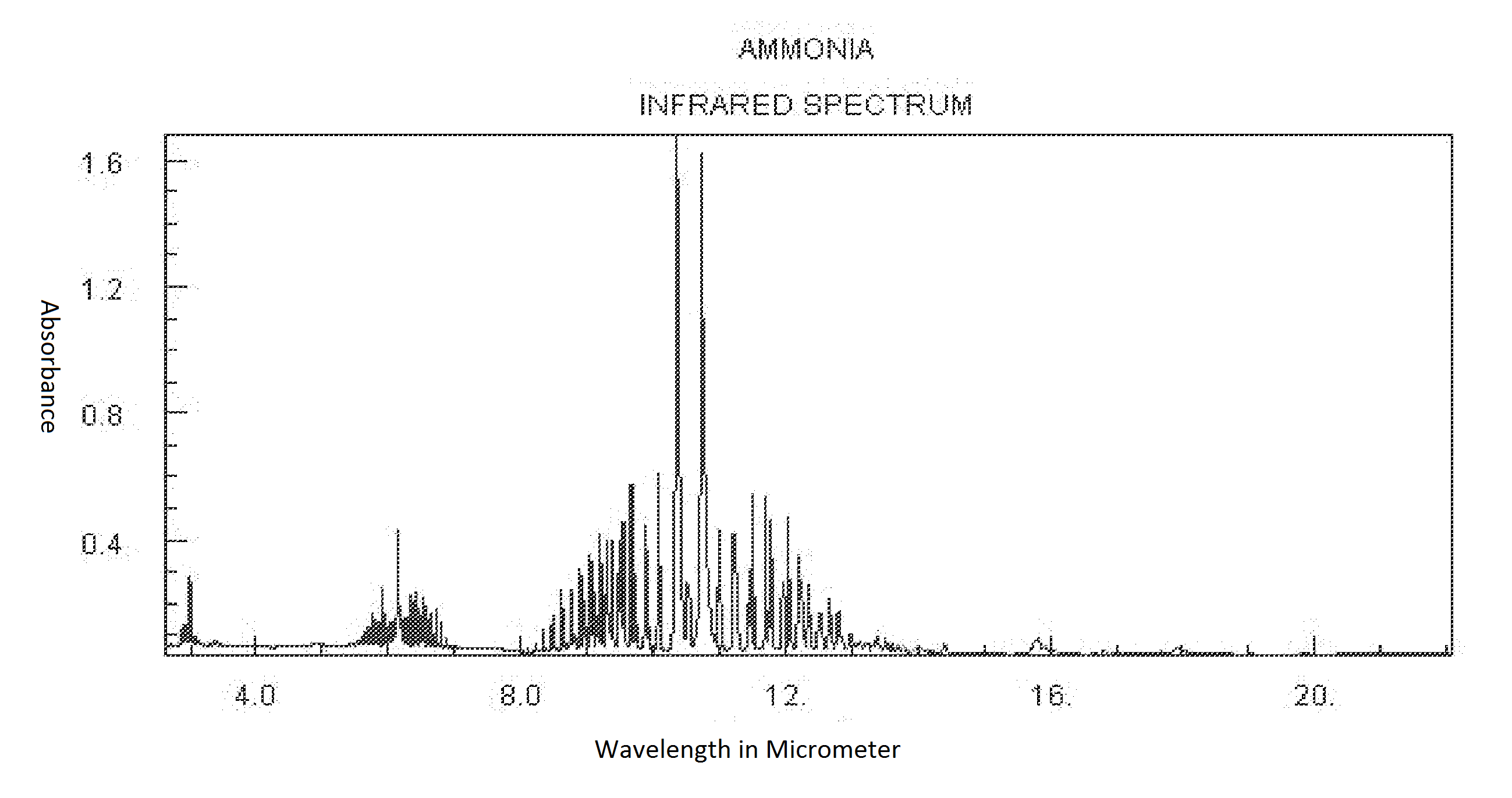}
    \caption{Infrared spectrum of ammonia. The figure is taken from NIST \cite{nist}.}
    \label{fig:ammonia_absorbance}
\end{figure}

\begin{figure}[b]
    \centering
    \begin{minipage}{.4\linewidth}
    \includegraphics[width=\linewidth, height=0.62\linewidth]{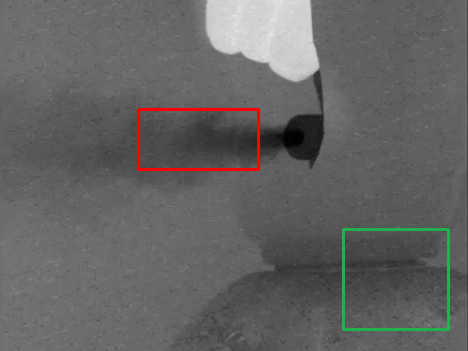}
    \end{minipage}
    \begin{minipage}{.4\linewidth}
    \includegraphics[width=\linewidth]{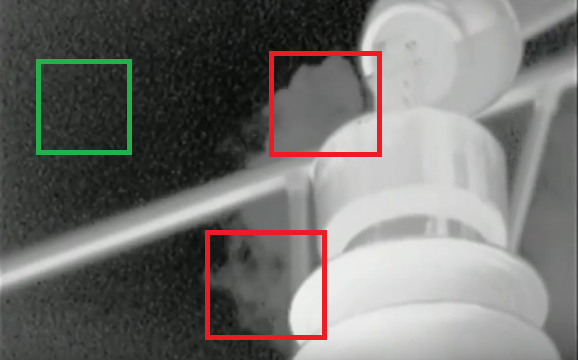}
    \end{minipage}
    \caption{Two infrared images of VOC gas leaks.
    %Green  rectangle corresponds to a leak-free region. 
    Red rectangles contain gas-leak regions. Green rectangles contain leak-free regions. Images are downloaded from FLIR systems and Infrared Cameras Inc. \cite{ferret.com.au_2012, cam_inc}, respectively.}.
    \label{fig:snapshot}
\end{figure}

%To prevent and detect gaseous leaks of harmful VOCs, we propose to develop a cyber-physical system (CPS) comprised of ordinary people or first responders using small, low-cost chemical sensors coupled to smart phones or mobile devices with the necessary computation and communication capabilities. Chemical sensors sample open air in a continuous manner and alert the first responders whenever they detect VOC and ammonia gas vapor leaks. Our goal is to detect VOC gas leaks and other dangerous high gas vapor concentration levels using un-calibrated low-cost sensors attached to smart mobile devices.
Some VOC gas vapors such as ethane and ammonia absorb infrared light in the Long Wave Infrared (LWIR) while others such as methane in the Medium Wave Infra-red (MWIR) bands. Absorbance by ammonia of infra-red light at different wavelengths is shown in Fig \ref{fig:ammonia_absorbance}. We can easily observe the existence of VOC gas vapor using  Infra-red (IR) cameras in open air as shown in Fig. \ref{fig:snapshot}. In this figure, a dark smoke-like region denotes the image of VOC gas vapor. However, the distance between the sensor and the source, and infrared reflections from the background significantly affect the recorded level \cite{erden2010voc,cetin2013method}.

Conventional optical devices, such as gas chromatographs and MWIR cameras, are generally expensive. A cheaper alternative would be the use of IR sensors and chemical gas sensors. Yet chemical gas sensors incur degradation in their sensitivity over time. Consequently, identically manufactured sensors are likely to yield significantly different responses upon exposure to gas analytes under identical conditions \cite{gopel1991definitions,davide1996frequency,zuppa2004drift,vergara2012chemical,artursson2000drift}. This problem is known in the literature as the \emph{sensor drift problem}.
%Sensor drift is mainly attributed to two types of reasons, namely, the physical changes in the structure of the sensor and the operating environment. The so-called \emph{first-order  drift} stems from aging and ``poisoning", an irreversible binding due to external contamination. The so-called \emph{second-order drift} is due to external and uncontrollable parameters such as temperature and humidity variations in the environment. As a result, it may not be a good idea to set a constant threshold to detect the existence of a gas leak or other dangerous substances using low-cost chemical sensors. Such sensors are additionally impacted by changes in the concentration of the gas in open air due to wind.

Causes of sensor drift can be summed up by two phenomena, namely, the physical changes in the structure of the sensor and the changes in the operating environment. The former case is known as \emph{first-order sensor drift}. It is caused by sensor aging or by sensor ``poisoning"\footnote{A process by which the sensor surface absorbs some compounds irreversibly, thus reducing its resistance sensitivity \cite{williams1995detection}.}. Unfortunately, neither poisoning nor aging are reversible as the physical structure of the sensor will have been permanently damaged or at the least affected. The latter case is known as \emph{second-order sensor drift} and is caused by external uncontrollable environmental changes, such as temperature and humidity variations. In this regard, the sensor response will be different from that expected from the original settings. Consequently, any decision thresholds that are optimal prior to sensor drift are likely to exhibit sub-optimal sensitivity and/or specificity once the aforementioned changes take place.  
%Chemical gas sensors constitute a potentially low-cost and practical alternative to conventional but expensive optical devices such as gas chromatographs and Medium Wave InfraRed (MWIR) cameras. However, lack of stability of measurements (which is called \emph{sensor drift}) is an important problem in chemical sensing. Selectivity and sensitivity of chemical sensors decrease over time. As a result, identically manufactured chemical sensors may generate significantly different results when exposed to the same analyte under identical conditions \cite{gopel1991definitions,davide1996frequency,zuppa2004drift,vergara2012chemical,artursson2000drift}.

\begin{figure}[t]
    \centering
    \includegraphics[width=\linewidth]{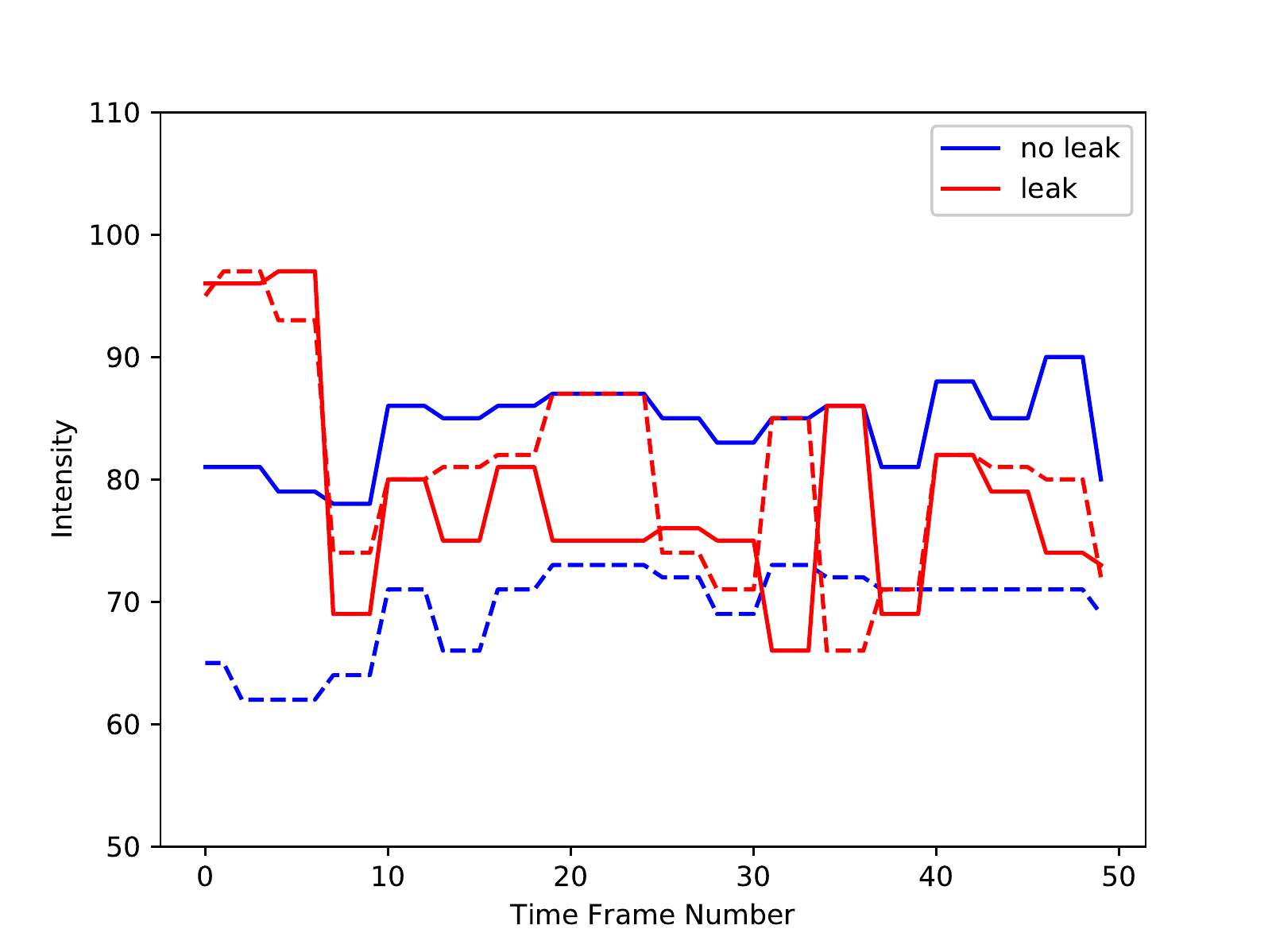}
    \caption{Example infrared sensor time-series data  near a gas leak. As one can see, it is not possible to find a threshold that can isolate leak-free intensity signals from those corresponding to a gas leak. Furthermore, the intensity of leak-free point signals is time-varying. This demonstrates the effect of noise, resolution and lighting factors that in turn lend further complexity to the task of distinguishing between the two classes of signals. The two sets of examples are distinguished by dotting. }
    \label{fig:voc_example}
\end{figure}

%Some VOC compounds and ammonia absorb infrared light at Medium Wave InfraRed (MWIR) and Long Wave InfraRed (LWIR) bands. 
%LWIR sensors are also able to detect ammonia gas leaks in open air.
%We can observe the existence of VOC gas vapor using  InfraRed (IR) cameras as shown in Fig. \ref{fig:snapshot}.However, the distance between the sensor and the source, and infrared reflections from the background significantly affect the recorded level \cite{erden2010voc,cetin2013method}. 
Similarly, while
 it is not possible to detect the concentration of the gas using MWIR and LWIR sensors in open air, it is possible to record a time-varying signal and detect the existence of gas leakage using IR sensors as shown in Fig. \ref{fig:voc_example} using a machine learning algorithm such as a neural network. The sensor signal exhibits sudden jumps and fluctuations due to gas vapor leak. Uncalibrated IR sensor intensity  measurements suddenly drop from 95 to 70 and fluctuate because of wind as shown in Fig. \ref{fig:voc_example}.
 
 {}
 
In this paper, we analyze the temporal sensor signals using convolutional, additive neural networks and the discriminator of a generative adversarial network (GAN) to detect and classify VOC gas leaks and other dangerous gas emissions. The proposed analysis is applicable to both Chemically-sensitive Field Effect Transistors (ChemFETs) and Electrochemical Impedance Spectroscopy (EIS) and infra-red sensors as they all produce time-varying signals.

%\vspace{-0.1cm}

The rest of the paper is organized as follows. Section \ref{sec:algo} describes the machine learning algorithms used in this paper. Section \ref{sec:data_and_meth} presents experimental results. We use an infrared data set and two publicly available chemical sensor drift data sets  obtained at the University of California at San Diego (UCSD) \cite{vergara2012chemical} and \cite{fonollosa2015reservoir}.
The paper finishes by offering a brief set of conclusions in Section \ref{sec:conclusion}.
%\vspace{-0.1cm}
\section{Deep Learning Algorithms for IR and Chemical Sensor Data Processing}
\label{sec:algo}
%\vspace{-0.1cm}
In this paper, we consider three tasks. Task 1 is infra-red sensor-based gas-leakage detection. In tasks 2 and 3, we identify different types of gas analytes.
%For each of our classification tasks, we use two novel neural network algorithms, alongside baseline conventional neural networks, in order to detect gas leakage (task 1), identify gas analytes (task 2 and task 2).

Our first network is an energy-efficient network, namely, an additive neural network, which is a neural network that performs no vector-multiplication except in its last layer. Our second neural network is the discriminator of a generative adversarial neural network, to which we refer shortly as DiscGAN. 
%We utilize three kinds of deep neural networks to classify the VOC gas data. The first one is a regular convolutional neural network (CNN). The second one is an additive neural network based on addition and sign operations, and the third one is the discriminator or a Generative Adversarial Neural (GAN) network.
%\vspace{-0.3cm}
\subsection{Convolutional Neural Networks}
\label{subsec:ConvNet}
%\vspace{-0.3cm}
Convolutional neural networks (or ConvNets) have been extensively used in computer vision \cite{krizhevsky2012imagenet,lecun1995convolutional} and time-series data analysis \cite{langkvist2014review}. In ConvNets, convolutions (or local correlations) between the inputs and the filter weights are used to extract local features at different scales in subsequent layers. %This locality in feature extraction, along with affine transformation and non-linearity, enables the network to extract highly abstract features at high levels.

%Since both the IR and chemical sensor readings are temporarily varying, we implemented a one-dimensional convolutional neural network with three convolutional layers including one dense input layer, one dense output layer and a binary classifier layer. The number of filters in the three filters are 16, 32 and 64, respectively. The size of each filter is 3 with maxpooling of size 5 being used between the layers. The output of the last maxpooling is subsequently flattened and fed to a dense layer of output size of 64, whose output is eventually fed to the binary classifier layer. We used rectified linear units (ReLU) as nonlinear activation in all hidden layers.

%\vspace{-0.3cm}
\subsection{Additive Neural Networks (AddNet)}
\label{subsec:AddNet}
%\vspace{-0.3cm}
Despite their ability to learn and recognize images and signals, deep learning algorithms are computationally expensive. This is attributed to the large number of add-multiply operations needed to be implemented in order to realize convolutions. This poses a problem when it comes to using such methods on platforms where energy is limited. As a result, simpler and, thus, more efficient algorithms are generally required to implement computationally expensive deep learning algorithms in such cases. 

Nevertheless, there have been attempts to leverage convolutional neural networks across energy-limited devices by means of methods that aim to either implement fewer dot-product operations, or to replace dot-product operations with computationally simpler operations.
%An example of the former is MobileNets \cite{howard2017mobilenets}, in which an expensive 2-dimensional convolution is replaced by the so-called \emph{separable convolution}. AddNets can be designed in a separable manner.
Binarizing the weights and/or the activations results in replacing real-number multiplication operations with binary logical operations when realizing convolution, as in the case of BinaryConnect\cite{courbariaux2015binaryconnect}, XNOR-Net\cite{rastegari2016xnor} and Binarized Neural Networks \cite{courbariaux2016binarized}.  
%success of deep learning algorithms in time-series analysis, CNNs are computationally expensive. It may not be possible to implement a CNN in regular mobile devices. Since multiplication operations consume significant energy, a regular neural network will fail to pass an energy efficiency constraint as well. 

An additive neural network (AddNet) falls under the second category, i.e., replacing real-valued multiplication operations in vector-vector and matrix-vector product operations by special addition operations. The new ``product'' operation comprises binary sign calculation, unsigned addition and regular addition.

In what follows, we define the scalar version of our binary operation and extend it straightforwardly to its vector operation. In this regard, let $x$ and $y \in \mathbf{R}$, the multiplication-devoid (abbreviated md) operation  denoted by $\oplus$ and defined as follows:
\begin{equation}\label{eqn_mf_def1}
    x \oplus y := \text{sgn}(x.y) (|x| + |y|)
\end{equation}
where sgn denotes the {\em signum} function. Alternatively, we can express the $\oplus$ operation as follows:
\begin{equation}
x \oplus y := \text{sgn}(x)y + \text{sgn}(y)x
%\vspace{-0.1cm}
\end{equation}
This is because $x=\text{sgn}(x)|x|$. One key property of the md operation is that it preserves the sign of regular multiplication operations \cite{badawi2017multiplication, afrasiyabi2018non}. We define the vector version of the md operation as follows. Let $\bf{x}$ and $\bf{w}$ be two vectors in $\bf{R}^{N}$. The md dot ``product'' is defined as:
%\vspace{-0.19cm}
	\begin{equation}
		\label{eqn_mf_vec_def}
		\mathbf{w}^T\oplus \mathbf{x} := \sum_{i=1}^N \text{sgn}(x_i.w_i ) (|x_i|+ |w_i|)
	\end{equation}

 It can be seen that the md operation expressed in Eq. \ref{eqn_mf_vec_def} requires no real-valued multiplication whatsoever. As such, instead of using add-multiply operations as in an ordinary dot product, we use ordinary addition and addition with sign multiplication in the md vector operation. Furthermore, we can restrict the operands $x_i$ and $w_i$ to be 8-bit numbers in order to speed up the vector addition operations. 
Another property of the md operation is that it induces the $\ell_1$ norm. This is shown as follows:
\begin{equation}
    \mathbf{x}^T \oplus \mathbf{x} = \sum_{i=1}^N \text{sgn}(x_i.x_i) (|x_i| + |x_i|) = 2{||\mathbf{x}||}_1
\end{equation}
In the context of neural networks, we use convolution and matrix-vector multiplication operations in convolutional and dense layers, respectively. In AddNet, we replace the aforementioned dot-product operations with the md vector product. The feed-forwarding pass in dense layers in a neural network can be expressed as follows:
\begin{equation}
o^l_i = \phi\big({\mathbf{w}_i^l}^T \mathbf{o}^{l-1} + b^l_i \big)
\end{equation}
where the superscript denotes the layer index, the subscript the neuron index, $\mathbf{w}_i^l$ the weights connecting the output of the previous layer (the $(l-1)$st layer) to the $ith$ neuron, $o_i^l$  the output of the $ith$ neuron, and bold $\mathbf{o}^{l-1}$ the vector output of the previous layer. $\phi$ is the non-linearity function applied element-wise and, finally, $b^l_i$ denotes the bias term added to the pre-activated response ${\mathbf{w}_i^l}^T \mathbf{o}^{l-1}$. 
%\newline
Similarly, we can define AddNet layers by replacing the dot-product ${\mathbf{w}_i^l}^T \mathbf{o}^{l-1}$ by our md operator as follows:
\begin{equation}\label{eqn:mf_nn_1}
o^l_i = \phi\big({\mathbf{w}_i^l}^T \oplus \mathbf{o}^{l-1} + b_i^l \big)    
\end{equation}

Since the md operator is additive, it will result in a larger output than ordinary multiplication does when either of the operands is of small magnitude, e.g. $3\oplus0.1=3.1> 3\times 0.1=0.3$. In the context of neural networks, the layer outputs and the weights are usually small values. As a result, the responses of the md layers will be of larger variance than those of the regular layer. This poses a problem in deep layers, where the dimension of the dot-product is quite large. In other words, if the depth of a convolutional layer is 64 and the kernel size is $3 \times 3$, the convolution operations will carry out dot-products between two vectors, each of which $\in \mathbb{R}^{3\times3\times64}$. In the case of the md layer, this will cause the output to exhibit inordinately high magnitudes. In order to overcome this, we introduce a scaling factor $\alpha$. As such, the feedforwarding pass in Eq. \ref{eqn:mf_nn_1} becomes
\begin{equation}\label{eqn:mf_nn_2}
o^l_i = \phi\big(\alpha_i^l({\mathbf{w}_i^l}^T \oplus \mathbf{o}^{l-1}) + b_i^l \big)     
\end{equation}

The scaling factor $\alpha^l_i$ enables us to control the range of the output prior to applying the activation function $\phi$ and, thus, leads to a controlled range of responses in subsequent layers. Note that the scaling by $\alpha^l_i$ in Eq. \ref{eqn:mf_nn_2} implies real-valued multiplication. Nevertheless, it requires only one real-valued multiplication per neuron. Therefore, carrying out scaling is not computationally expensive. Numerous options exist for selecting the scaling factor $\alpha^l_i$. One possibility may be the setting of $\alpha^l_i$ to $\frac{1}{{||w_i^l||}_1}$, i.e., the reciprocal of the $\ell_1$ norm of the associated weights. Another option would be having $\alpha^l_i$ be trainable by backpropagation. The latter delivers more flexibility for the model.

Nevertheless, batch normalization is a common practice in neural networks and has shown to be quite effective in accelerating the training of deep networks \cite{ioffe2015batch}. Therefore, one can simply apply batch normalization to the pre-activation responses in AddNet. Such normalization eliminates the need to carry out scaling by $\alpha^l_i$ as it will be subsumed by the scaling induced by batch normalization.

The proof of AddNet with linear and/or ReLU activation functions satisfying the {\em universal approximation property} over the space of Lebesgue integrable functions can be found in  \cite{cybenko1989approximation}.

\begin{figure}[t]
    \centering
    \includegraphics[width=\linewidth]{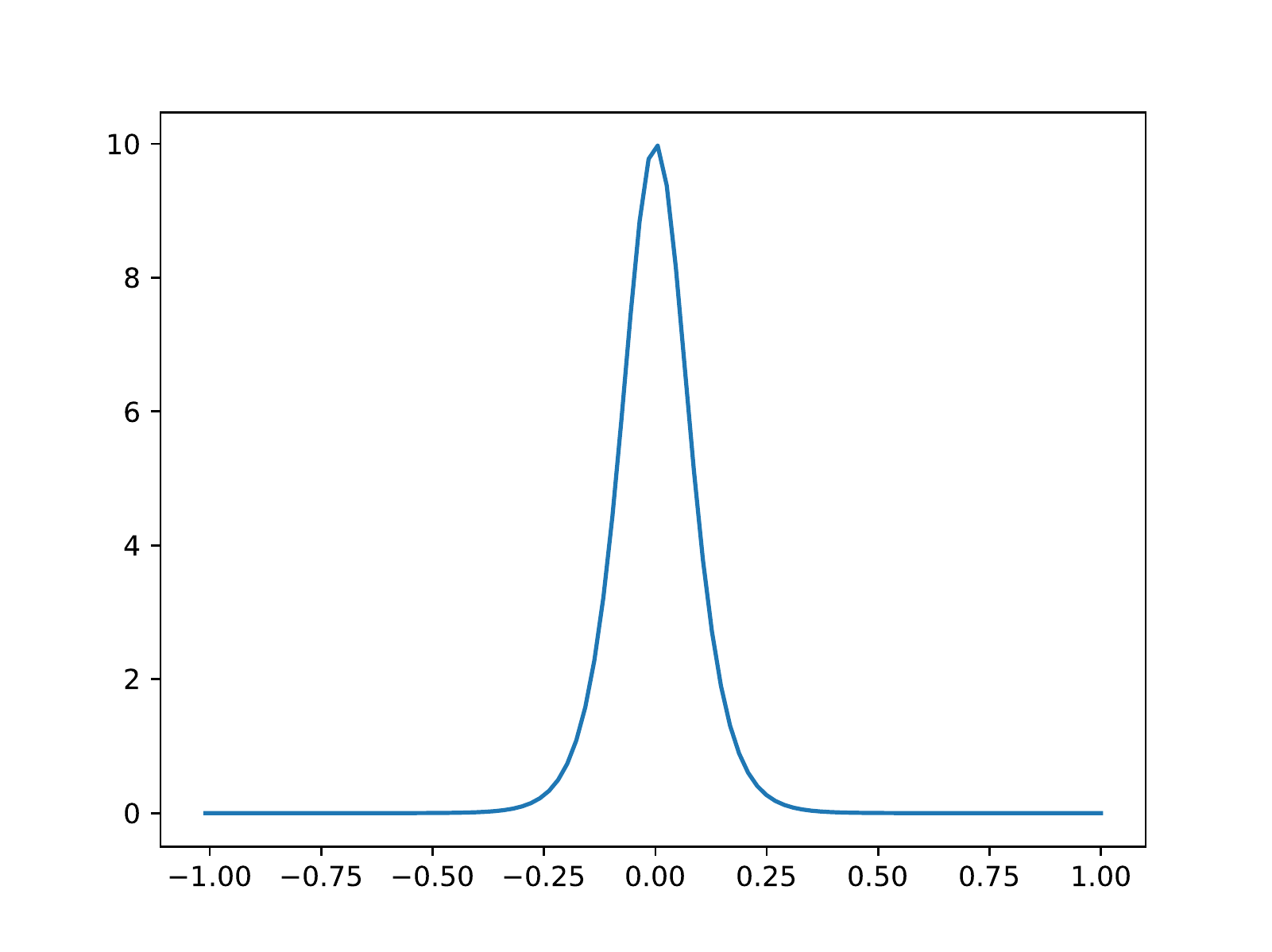}
    \caption{The derivative of $\tanh(a w) = a \sech^2(a w)$ as a function of parameter $w$,  with $a$ set to 10.}
    \label{fig:sech_sq}
\end{figure}

As for training the md layers by backpropagation, it is worth mentioning that the derivative of the {\em signum} function used in the definition has to be computed. This is because $\frac{d ~ \text{sgn}(w)}{dw} = 2 \delta(w)$, where $\delta$ is the Dirac-delta function. 
In practice, this means that the derivative of the {\em signum} function is zero almost everywhere except when $w=0$. 
%This results in the gradient flowing from one term of the md-operator, as the partial derivative will be (in the scalar case).
%We need to compute 
The partial derivative of the md operator w.r.t. $w$ is:
\begin{equation}
    \frac{\partial (w \oplus x)}{\partial w} = \text{sgn}(x)+ 2x\delta(w)
\end{equation}
We approximate the derivative of the {\em signum} operator using the hyperbolic tangent as follows:
\begin{equation}\label{eqn:mf_operator_derv2}
    \frac{d ~ \text{sgn}(w)}{d w} \approx \frac{d ~ \tanh(aw)}{d w} = a \sech^2(aw)
\end{equation}
where $\sech(x) = \frac{2}{e^x+e^{-x}}$ is the hyperbolic secant function, and $a$ is a hyperparameter indicating how sharp the hyperbolic tangent is. The larger the hyperparameter $a$ is, the closer $\tanh$ is to the {\em signum} function. Figure \ref{fig:sech_sq} shows the approximate derivative of the {\em signum} function for $a=10$.
As we can see from Fig. \ref{fig:sech_sq}, the derivative has high magnitude for $w$ values close to zero, whereas it is effectively zero for large values. This can be seen as allowing small weights to have finer updates than larger weights and thus allowing them to change their sign more often during training.
We found empirically that this approximate derivative computation provides satisfactory convergence rates in Google's Tensorflow software.

%we redefined the derivative of the {\em signum} function in the operator to be that of the hyperbolic tangent function. Since the hyperbolic tangent function acts linearly close to zero, this makes the derivative smooth for small values. 
%This leads to fine gradient updates during network training. 
%This showed to improve the accuracy of the network and accelerate convergence. 
%\vspace{-0.3cm}
\subsection{DiscGan (Discriminator of GAN as Classifier)}
%\vspace{-0.3cm}
Generative Adversarial Networks (GAN) have become the benchmark in image synthesis \cite{goodfellow2014generative, radford2015unsupervised}. A typical GAN has a generative network, which attempts to generate images (or data) resembling real images from noise input, and a discriminator network, which attempts to discriminate between the real images and those synthesized by the generator. The generator and the discriminator are optimized in an adversarial scheme, i.e., the generator tries to fool the discriminator by the synthetic data it produces, and, in turn, the discriminator tries to counteract the generator by discriminating between the real data samples and the fake ones. 

In this paper, our aim is not to synthesize realistic data but rather to make use of the adversarial nature of GAN training in order to obtain a discriminator network capable of classifying the input with an unbalanced set of training data. As the recordings of gas leak data may fall short of the clean air recordings for this purpose, we have the generator of the GAN to compensate for the data set with a smaller number of data instances by producing ``artificial" gas leak data during training.

In this regard, we perform a two-phase training of the GAN. First, we carry out adversarial training of both the discriminator and the generator using the data of one of the classes. In the second phase, we use data from both classes and carry supervised binary-classification training of the discriminator which now acts as a classifier.
In this setting, let $x^i$ represent the $ith$ data instance of one of the classes.  In this case, $x^i$s denote the gas leak recordings (or the anomalous class). Let $z$ be a random noise vector, e.g. Gaussian noise or uniform noise. Let $D$ be the discriminator and $G$ be the generator, with each having a set of parameters $\theta_D$ and $\theta_G$, respectively. In the adversarial-training phase, we seek to optimize the following loss function:
%\vspace{-0.1cm}
\begin{equation}\label{eqn:GAN_1}
\max_{\theta_D} \min_{\theta_G} \sum_i \log(D(x^i)) + \sum_i \log(1-D(G(z^i))
%\vspace{-0.1cm}
\end{equation}
where %$x^i$ is a real data point of the class of interest. 
$D(x^i)$ is the soft prediction result of the discriminator corresponding to data point $x^i$. From the discriminator perspective, the prediction output $D(x^i)$ should be close to 1 because $x^i$ is ``real". The generator $G$ produces ``fake" data signals from noise vector $z^i$, that is $G(z^i)$, and the prediction $D(G(z^i))$ should be close to zero because $G(z^i)$  is an artificial data instance. The generator, on the other hand, will try to produce $G(z^i)$ that will be assigned the prediction $D(G(z^i))$ close to 1. Once training the first stage is accomplished, we move on to the second stage of supervised training of the entire training data, in which the cost function we seek to minimize is the regular binary cross entropy function $CE$ expressed as follows:
%\vspace{-0.1cm}
\begin{equation}\label{eqn:cross_entropy}
%\begin{split}
   CE :=   -\frac{1}{N} \Big( \sum_i (1-t^i)\log(1-D(x^i))%\\&
   + t^i \log(D(x^i)) \Big)
%   \end{split}
%\vspace{-0.1cm}
\end{equation}
where $t^i ~ \in\{0,1\}$ denotes the true class of $x^i $.

When there are multiple classes we can still use
the discriminator of a GAN with a slight modification of the loss functions. In this regard, let us assume that there are $N$-classes. In this case, the one-hot encoded label for each input is an $N$-dimensional vector, with all entries equal to zero, except for the $kth$ entry, where $k$ is the true class. During training, the discriminator (or classifier) will minimize the cross entropy of the softmax layer applied at the output layer ($N$ logits). The generator $G$ will attack the output of the $kth$ node. Here we consider the output of the $kth$ node to be the logit of a binary class, i.e., the adversarial loss criterion becomes:
\begin{equation}\label{eqn:ce_g}
    max_{\theta_D} min_{\theta_G}\log(D_{k}\big(G(z)\big) 
 %   \vspace{-0.1cm}
\end{equation}
%The generator parameters $\theta_G$ will try to attack the $k^{th}$ output node by trying to push the sigmoidal response of that node to unity by minimizing the cross entropy $CE_G$ of that node, which is expressed as:
%\vspace{-0.1cm}
%\begin{equation}\label{eqn:ce_g}
%    CE_G(node\ k) = -\log(D_{k}(G(z)) 
% %   \vspace{-0.1cm}
%\end{equation}
where $D_{k}\big(G(z)\big)$ is the discriminator \emph{sigmoidal} response of the $kth$ node, i.e., we apply the sigmoid function to the logits before taking the logarithm in determining the loss.
%The discriminator parameters $\theta_D$ in turn will counteract the generator by trying to push the sigmoidal response of the fake examples to zero by minimizing the cross entropy $CE_D$ given by
%\vspace{-0.1cm}
%\begin{equation}\label{eqn:ce_d}
%    CE_D(node\ k) = -\log(1-D(G(z))
 %   \vspace{-0.1cm}
%\end{equation}
Note that the loss here is different from the multi-class case, in which we consider multi-class logits, i.e., we use sigmoid normalization instead of the softmax normalization. %as we reduce the problem into a two-class adversarial optimization. Therefore, for each data point, only one output node will be attacked at a time. 
%which has the location of the true class. 
In practice, since we do a mini-batch update, we take the average of the loss functions and minimize the loss functions based on the mini-batch gradients. 
%Note that since the discriminator will be driven to assign a probability of 0 to "true" images, one should preserve consistency by assigning true labels of 1 to the class data points used in stage 1, and a true label of 0 to the otherwise class.
%\vspace{-0.5cm}
\section{Datasets and Experimental Results}
\label{sec:data_and_meth}
%\vspace{-0.5cm}
\subsection{Infra-red VOC Dataset}
\label{ssec:voc_dataset}
%\vspace{-0.3cm}
Our first data set consists of infra-red imaging signals of VOC gas leaks in open air and clean air recordings. Specifically,  we have two classes of discrete-time signals corresponding to VOC gas leaks and clean air, respectively. Each signal is a time series containing 50 samples corresponding to two seconds of recording with a sampling rate of 25 samples per second. The recorded value varies in open air because of background temperature variations and low resolution error as it can be observed in Fig. \ref{fig:voc_example}. Furthermore, the sensors may not be calibrated in practice, so their sensitivity may differ across time. % When there is no leak, the variance is much smaller compared to the VOC gas leak in which case the gas molecules absorb the IR light.
We gathered about 30,000 VOC gas leak and 30,000 clean air data instances.%The dynamic range of the recorded signal values varies according to the background temperature and the sensors are uncalibrated in practice as shown in Fig. \ref{fig:voc_example} Therefore,

%An infrared image of the gas leak is shown in Fig. \ref{fig:snapshot}.
The images are obtained using an MWIR camera produced by FLIR systems and Infrared Cameras Inc. \cite{ferret.com.au_2012,cam_inc}. VOC gas absorbs the infra-red light appearing as a white cloud in the black-hot mode infra-red image as shown in Fig. \ref{fig:snapshot}. In these videos, a gas leak erupts from the source with the gas spreading out as time progresses. We manually selected regions of interest and assigned normal event designations to temporal measurements where no gas is present throughout these series, while designating the rest as anomalous events. 

We used  min-max normalization in order to scale signal data points between 0 and 1. The normalized signal $\hat{x}$ is obtained as follows
%The normalization is expressed in the following equation:
%\vspace{-0.1cm}
\begin{equation}
\label{eqn:minmax}
    \hat{x}[n]=\frac{x[n]-\min(x)}{\max(x)-\min(x)},\quad n=0,1,... \ ,49
 %   \vspace{-0.1cm}
\end{equation}
where $\max(x)$ and $\min(x)$ represent the maximum and minimum values of a given infrared signal $x$, respectively. 
%the average of a $3 \times 3$ area of interest for $50$ consecutive frames.     

%We implemented  three algorithms described in Sec.\ref{sec:algo}. 
We used convolutional neural networks with the architecture specified in Table \ref{tab:cnn_data1_arch}. In order to obtain more temporal data points, and in order to ensure the network is translation-invariant to the gas eruption location, we chose to randomly crop the input data into temporal signals of size $32$ each.
\begin{table}[b]
    \centering
        \caption{Architecture of the convolutional neural network used in classifying the data set of Sec. \ref{ssec:voc_dataset}}
    \begin{tabular}{c|c}
       \toprule
        Layer &  Specification\\
        \midrule
         Input Layer& input size: $32 \times 1$\\ 
         Conv Layer & 16 $3\times 1$ filters, no strides\\
         Max-pooling Layer & down-sampling by 2\\
         Batch-normalization Layer& -\\
         \midrule
         Conv Layer & 32 $3 \times 16$ filters, no strides\\
         Max-pooling Layer & down-sampling by 2\\
         Batch-normalization Layer& -\\
         \midrule         
         Conv Layer & 64 $3 \times 32$ filters, no strides\\
         Max-pooling Layer & down-sampling by 2\\
         Batch-normalization Layer& -\\
         \midrule
         Global Average-pooling Layer&output size: 64\\
                  \midrule         
         Dense Layer& output size: 64\\
         Batch-normalization Layer&-\\
         Output Linear Layer& output size: 1\\
         \bottomrule
    \end{tabular}

    \label{tab:cnn_data1_arch}
\end{table}
We divided our data set into three disjoint sets. The training data consists of 8,000 recordings of each class. Another set of 8,000 recordings of each class is used as the validation data set. The rest of the data was reserved for testing. We trained our networks using the RMSProp optimizer algorithm \cite{tieleman2012lecture}. We tested the hypothesis of whether dropout helps achieve better results \cite{srivastava2014dropout} by using a dropout rate of $50\%$. As for the GAN approach, we used a generator which is a multi-layer perceptron (MLP) with one hidden layer of size 256. 
%Classification results are summarized in Table \ref{Tab:VOC_results}. 
The regular convolutional neural network and AddNet exhibit comparable results.  We obtained an accuracy of $99.8\%$ for no-gas data and $99.7\%$ for gas-leak data for a regular ConvNet. AddNet attained a recognition rate of $98.9\%$ for no-gas data and $99.3\%$ for gas-leak data.
%The best classification accuracy is obtained using the AddNet with "no dropout" case.

In the second set of experiments, we assumed that we have an unbalanced data set. In practice, we may not have VOC or ammonia gas leak recordings as clean air. We trained the models with only 50 recordings of gas leak signals against 8,000 recordings of clean air recordings. The test data set contains 14,000 recording instances of VOC gas leaks and clean air recordings. Classification results are also summarized in Table \ref{Tab:VOC_results}. AddNet produces the best results but the discriminator of the GAN Network is also quite close to AddNet. The confusion matrix of the results of the best model is given in Table \ref{tab:conf_mat_data1}.

\begin{table}[t]
 \centering
   \caption{Accuracy results for infra-red VOC data. Classifiers are trained with only 50 VOC gas leak recordings vs 8000 clean air recordings.}
  \begin{tabular}{c|c|c|c}
     \toprule
      Model & \begin{tabular}{@{}c@{}}No-gas \\Accuracy\\(specificity)\end{tabular} &
      \begin{tabular}{@{}c@{}}
             Gas-leak\\
             Accuracy\\
             (sensitivity)
      \end{tabular} &\begin{tabular}{@{}c@{}}
           Total \\
           Accuracy
      \end{tabular}\\ \midrule
        \begin{tabular}{@{}c@{}} ConvNet\\(dropout $50\%$) \end{tabular}& $98.3\%$ & $95.8\%$ & $97.1\%$\\ %\hline
        \begin{tabular}{@{}c@{}} ConvNet\\no dropout \end{tabular} & $98.0\%$ & $94.2\%$ & $96.1\%$\\ %\hline
       \begin{tabular}{@{}c@{}} AddNet\\(dropout $50\%$) \end{tabular} & $98.2\%$ & $96.0\%$ & $97.1\%$\\ %\hline
        \begin{tabular}{@{}c@{}} AddNet\\(no dropout) \end{tabular} & $99.1\%$ & $97.3\%$ & $98.2\%$\\ %\hline
       \begin{tabular}{@{}c@{}}
            DiscGAN\\
             
       \end{tabular}  & $99.0\%$ & $97.1\%$ & $98.1\%$\\ \bottomrule
  \end{tabular}

  \label{Tab:VOC_results}
\end{table}

\begin{table}[b]
    \centering
    \caption{Confusion matrix for the best achieving neural network (AddNet with no dropout) over the testing data. The true positive rate (sensitivity) is $97.3\%$. The true negative rate (specificity) is $99.1\%$. }
    \begin{tabular}{ccccc}
    \toprule
     \multirow{3}{*}{\begin{tabular}{@{}c@{}}Actual \\ Class\end{tabular}}&\multicolumn{2}{c}{\begin{tabular}{@{}c@{}}
      Predicted Class\\\bottomrule
 \end{tabular}}&\multirow{3}{*}{\begin{tabular}{@{}c@{}}
      Total \\Count
 \end{tabular}}\\
 \multicolumn{2}{c}{}Leak&No Leak&\\
 \multicolumn{2}{c}{}(positive)&(negative)&\\
 \midrule
         Leak (positive) &13,622
 &378&14,000\\
         No Leak (negative)&126&13,874 &14,000\\
         \bottomrule
    \end{tabular}
    \label{tab:conf_mat_data1}
\end{table}

We also investigated pruning the weights in both AddNet and ConvNet during inference. In this regard, we discard the magnitudes of the smallest magnitude weights while retaining their sign information. We keep the bias coefficients and the coefficients of the last layer intact. Results of various pruning rates are shown in Table \ref{tab:compression}. Apparently, in AddNet, we can discard the magnitude information of the weights up to a high rate ($67.4\%$) without severely degrading performance. On the other hand, the magnitude information is quite critical in the case of a regular ConvNet. These results clearly show the advantages of AddNet, which requires reduced memory space in a mobile device and consumes less energy as it performs much fewer arithmetic operations during inference.  
%\vspace{-0.2cm}
\begin{table}[!htbp]
    \centering
\caption{Effect of compressing weights of AddNet and ConvNet by discarding the smallest $K\%$ magnitude while keeping the sign information. ConvNet fails to produce reasonable results when the compression rate exceeds $16.1\%$. The compression rate is estimated by allocating 32 bits to intact weight values and 1 bit for every binarized weight factor.}
    \begin{tabular}{ccccccc}
     \toprule       
            \begin{tabular}{@{}c@{}}
                  Model\\
                  Accuracy
            \end{tabular} &
 \multicolumn{6}{c}{\begin{tabular}{@{}c@{}}
      Weight Compression\\ Rate (smallest K\%) \\
       \bottomrule
 \end{tabular}}\\
             
         & $0$ &$16.1$ & $19.7$ & $67.4$ & $76.8$& $86.6$ \\\toprule
            AddNet  & 98.9 &97.2&97.9&98.0&97.1& 61.3\\\midrule
         ConvNet  & 99.8 &67.4& $-$&$-$ &$-$&$-$\\
         \bottomrule
    \end{tabular}

    \label{tab:compression}
\end{table}

\subsection{Gas Sensor Array Recordings under Dynamic Gas Mixtures}\label{sec_data2}
We consider a gas type identification problem, in which we have three types of gases to identify, namely, CO, Ethylene and Methane. We used the data set obtained by Fonollosa et al. \cite{fonollosa2015reservoir}. The data set consists of time-series measurements of a sensor array of 16 chemical metal-oxide sensors under exposure to two different kinds of gas mixtures, ethylene and methane in air, and ethylene and CO in air. Sensors were exposed to volatile organic compounds at different concentration levels under tightly-controlled operating conditions during the experiment. The data is obtained at a sampling frequency of 100 $Hz$. The 16 chemical sensors are of four different types, with each having four identical sensors.\footnote{For more details, the reader may refer to the original paper \cite{fonollosa2015reservoir}.}
 Furthermore, switching between different mixtures of VOCs may occur too fast making it challenging if not impossible for the sensors to reach steady state. This makes identifying the gas analytes 
 difficult using a machine learning method.
 % inasmuch as the effect of the previous response takes place.   

\begin{figure*}[!htbp]
   \begin{minipage}{0.33\textwidth}
     \centering
     \includegraphics[width=0.95\linewidth,height=0.7\linewidth]{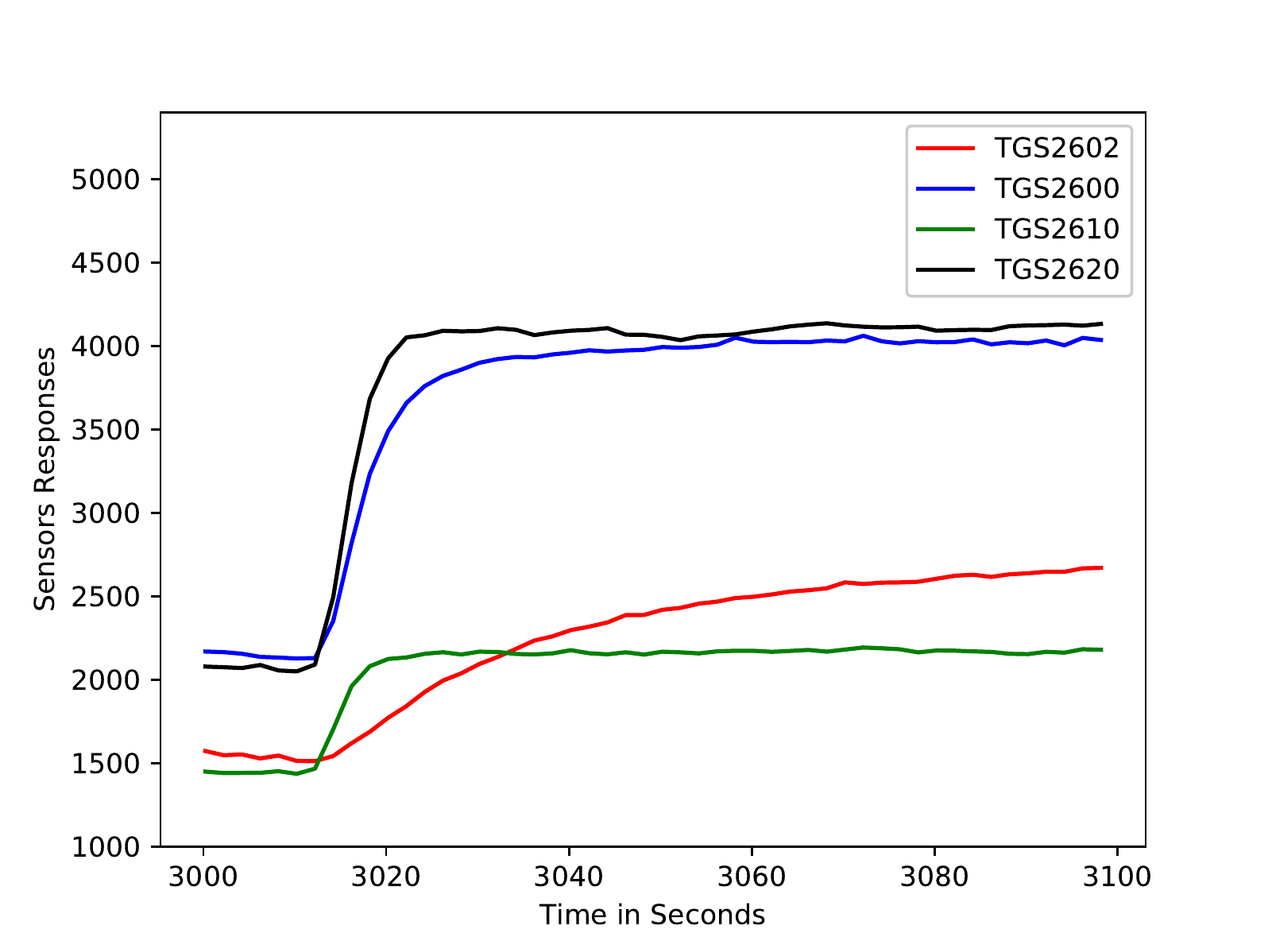}
     (a) CO
     %\caption{Time-series data generated by four different sensors under CO exposure. \newline}\label{Fig:co}
   \end{minipage}\hfill
   \begin{minipage}{0.33\textwidth}
     \centering
     \includegraphics[width=0.95\linewidth,height=0.7\linewidth]{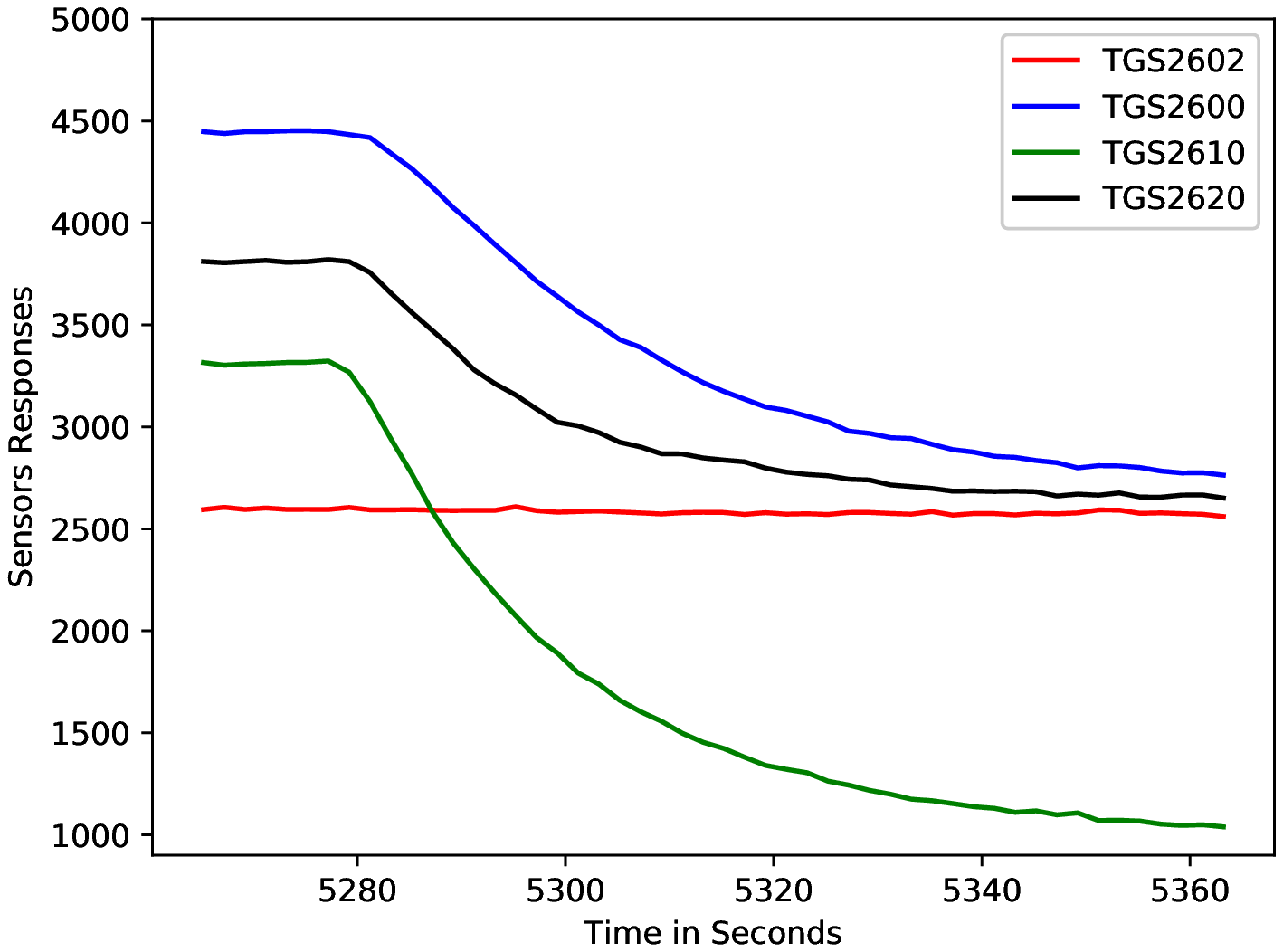}
     (b) Ethylene
     %\caption{Time-series data generated by four different sensors under ethylene exposure. }\label{Fig:eth}
   \end{minipage}\hfill
   \begin{minipage}{0.33\textwidth}
     \centering
     \includegraphics[width=0.95\linewidth,height=0.7\linewidth]{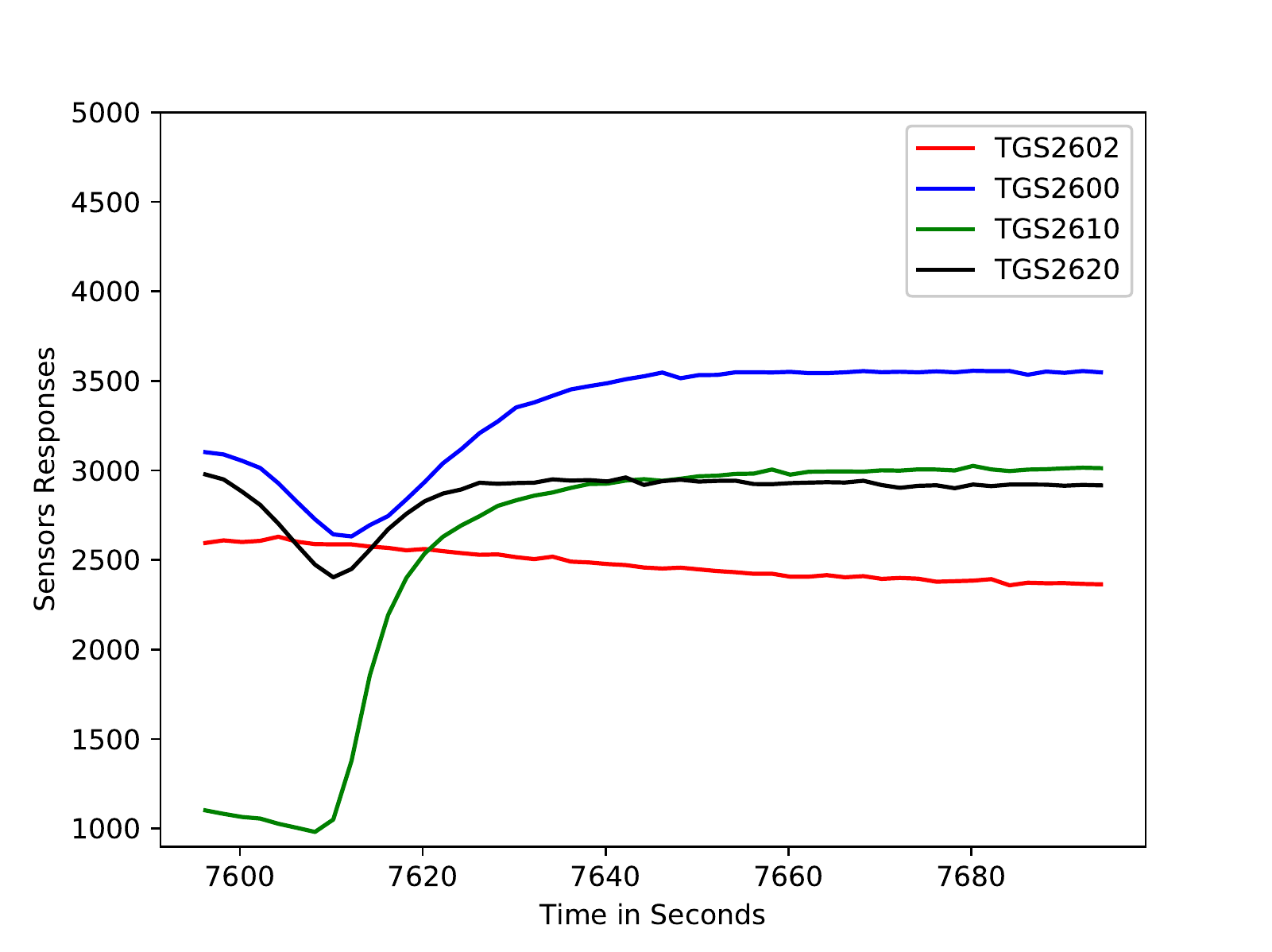}
     (c) Methane
     %\caption{Time-series data generated by four different sensors under methane exposure.}\label{Fig:methane}
   \end{minipage}\hfill
   \caption{Time-series data generated by four different sensors under exposure to different type of gases (50 time samples for each sensor).}
   \label{Fig:three_gases}
\end{figure*}

The recorded sensor data is deposited to the UC Irvine Machine Learning Repository online in the form of two long time-series.
We extracted portions of the time series such that the sensor array is exposed to one type of analyte at a given time. Each recording corresponds to 100 seconds of data. We observed that it is enough to sample the sensor response every 2 seconds.
%We down-sampled the UCSD data by a factor of 200. Thus we considered a sampling rate of 0.5 samples/second. 
Example sensor response signals to CO, ethylene and methane gas vapor exposures are shown in Fig. \ref{Fig:three_gases}. Each sub-figure contains four different sensor responses.

We gathered a total of 215 instances from the raw recordings, in which we have 49 CO, 116 ethylene and 50 methane time-series signals. Each instance has $50$ time measurements for each sensor. Thus, a total of $50 \times 16$ measurements per instance are used. Since the number of instances in the data set is small, we employed  cross validation with holdout method, where our validation set consists of 35 examples, with the experiment repeated 4 times. Thus, we validated our results over 140 examples. Furthermore, since the number of instances is small compared to the input dimension $50 \times 16$, we opted to randomly crop data points during training into smaller time series of size $40\times16$. This allows the classifier to be invariant to the exact time where the exposure takes place. Furthermore, it increases the number of data points during training.

%Since the sensors belong to different modality (different types), and since sensors of the same type have different temporal responses, we chose to process the data spatio-temporally by feeding it to 1-dimensional convolutional neural networks, with a spatial dimension of 40 and channel (depth) dimension of 16, each of which corresponding to the response of a different sensor. This allows the neural network to learn cross-sensor features while preserving the time-series structure. The architecture of our neural network is given in Table \ref{tab:cnn_data2_arch}.
Since the sensors are of different types, and since even the sensors of the same type produce different temporal responses, we process the temporal sensor data using 1-D convolutional networks. The input to each neural network is a matrix of size $40 \times 16$, for $40$ time instances and 16  sensors.

\begin{table}[t]
    \centering
    \caption{Architecture of the convolutional neural network used in classifying the data set of Sec. \ref{sec_data2}.}
    \begin{tabular}{c|c}
       \toprule
        Layer &  Specification\\
        \midrule
         Input Layer& input size: $40 \times 16$\\ 
         Conv Layer & 64 $5\times 16$ filters, no strides\\
         Max-pooling Layer & pooling size: 4, stride size: 3\\
         Batch-normalization Layer& -\\
         \midrule
         Conv Layer & 128 $5 \times 64$ filters, no strides\\
         Max-pooling Layer & pooling size: 4, stride size: 3\\
         Batch-normalization Layer& -\\
         \midrule         
         Conv Layer & 256 $5 \times 128$ filters, no strides\\
         Max-pooling Layer & pooling size: 4, stride size: 3\\
         Batch-normalization Layer& -\\
         \midrule
         Global Average-pooling Layer&output size: 256\\
                  \midrule         
         Dense Layer& output size: 256\\
         Batch-normalization Layer&-\\
         Output Linear Layer& output size: 3\\
         \bottomrule
    \end{tabular}

    \label{tab:cnn_data2_arch}
\end{table} 

We used ReLU non-linearity between layers. Our loss function is the cross-entropy with the softmax operator. We used the RMSProp optimizer to carry out the parameter updates during training. We trained a regular ConvNet and an AddNet of the same architecture as in Table \ref{tab:cnn_data2_arch}. Our classification accuracy results over the testing data are shown in Table \ref{tab:data2_res}. The confusion matrix of the results over the validation data set obtained by AddNet is given in Table \ref{tab:conf_matrix_data2}.
\begin{table}[b]
    \centering
    \caption{Recognition rates for the two neural networks over the test data set}
    \begin{tabular}{ccccc}
    \toprule
    \multirow{2}{*}{}
          &\multicolumn{3}{c}{\begin{tabular}{@{}c@{}}
      Gas Type-based Accuracy\\\bottomrule \addlinespace[0.1em]
 \end{tabular}}&\multirow{2}{*}{\begin{tabular}{@{}c@{}}
      Average \\
       Accuracy
 \end{tabular}}\\&CO &Ethylene &Methane&\\\toprule

         ConvNet &$91.1\%$ &$98.6\%$& $100\%$ &  $96.6\%$\\
         AddNet &$91.1\%$&$97.2\%$ & $100\%$ &$96.1\%$  \\ \bottomrule
    \end{tabular}

    \label{tab:data2_res}
\end{table}{}
\begin{table}[b]
\centering
\caption{Confusion matrix for AddNet over the validation data sets for repeated trials }
\begin{tabular}{ccccc}
   \toprule
     True&\multicolumn{3}{c}{\begin{tabular}{@{}c@{}}
      Predicted Class
 \end{tabular}}&Total\\
    Class&CO&Ethylene&Methane& Count\\\toprule
     CO&31&3&0&34\\
     Ethylene&2&70&0&72\\
     Methane&0&0&34&34\\
     \bottomrule
\end{tabular}
\label{tab:conf_matrix_data2}
\end{table}

It can be observed in Table \ref{tab:data2_res} that the recognition capabilities of both AddNet and ConvNet are on par with one another. It is worth emphasizing the computational frugality of the scheme as use of the regular dot-product is confined solely to the last layer in AddNet.

\subsection{Chemical Gas Sensor Array Drift Dataset}
%\vspace{-0.2cm}
The third data set is the publicly available chemical VOC gas sensor drift data set compiled by Vergara \textit{et al.} at UCSD \cite{vergara2012chemical}. The data set was obtained by exposing an array of 16 distinct chemical sensors to 6 types of gas mixtures (ammonia, acetone, ethylene, ethanol, toluene and acetaldehyde) at a variety of concentration levels. Each data record is  a vector time series. Vectors contain 8 feature parameters extracted from the sensor time series signals during a gas release experiment, conducted over a period of three years at UCSD. 
The feature parameters include the steady state resistance value and the normalized resistance change. The remaining 6 parameter features are the maxima and minima of the exponential moving average
(ema$_\alpha$) transform governed by the following input-output relation:
\begin{equation}
    y[k] = (1-\alpha)y[k-1] + \alpha(r[k]-r[k-1])
\end{equation}
where $r[k]$ is the resistance value at time step $k$, and $y[k]$ is the transformed value after applying the ema filter. The maxima and minima features are reported for $\alpha$ values equal to $0.1$, $0.01$ and $0.001$ over an entire experiment. These ema features have distinct time constants for different $\alpha$ values, as they contain temporal information.
%Six of the features of each sensor account for the minimum and the maximum values of the ema$_\alpha$ transform. %Furthermore, the maximum absolute response of the raw readouts and its normalization account for two more features.
Unfortunately, the raw time-domain sensor signals are not available in this data set.

Since there are 16 sensors, 
 a total of $16 \times 8 = 128$ feature values are recorded per experiment. 
%These experiments were carried out over a period of 36 months and data was collected  throughout. 
%The main problem tackled by the work focused on the investigation of the usage of machine learning so as to compensate for degradation in the sensor measurements due to sensor drift effect over time.
The data set is divided into 10 batches ordered chronologically. Full details about the experiment and the data set can be found in \cite{vergara2012chemical}.

We carried out our classification tasks by training neural networks for $N=2$ batches and testing on successive batches. This  is identical to the sensor drift estimation  approach given in \cite{vergara2012chemical}.
%in order to see how much compensation the neural network is capable of over a variety of sensor array drift periods.\parskip
Because feature values have huge variances, we opted to apply the signed square root function element-wise to control the ranges of the reported values. The modification delivered improved results in our experiments, especially for later batches.

We trained an MLP model with two hidden layers, each with 512 output units, and an output layer. Furthermore, we trained the network for 100 epochs using the RMSProp optimizer \cite{tieleman2012lecture}. We applied a dropout rate of $20\%$ and used a batch size of 128 in order to prevent complex co-adaptation. To augment the data, we added a zero-mean Gaussian noise with standard deviation of 0.1.

We also tried combining AddNet with the GAN approach, in which case, the discriminator is an AddNet and the generator is a regular MLP. The architecture of the network is the same as that of the GAN we use. Furthermore, we tried utilizing the other batches by passing them to the classifier and carrying out backpropagation according to their guessed labels. This is done in order for the network to utilize the correctly guessed labels so that it could be helpful in improving the classification accuracy for the mis-classified data point. A numerical comparison of the proposed methods to the SVM-classifier ensemble used in \cite{vergara2012chemical} is given in Table \ref{tab:uci_numerical_table}. In general, the AddNet-MLP,
the MLP and the multi-class GAN discriminator produce better sensor-drift compensated results than does the SVM based method.

\begin{table*}[!htbp]
\centering
\caption{Comparative accuracy (in $\%$) results of the various models when training on batches 1 and 2 and testing on batches 3-10. Bold-text numbers correspond to the best accuracy results obtained amongst the different algorithms for each batch.}
\begin{tabular}{cccccccc}
     \toprule
     Batch ID& \begin{tabular}{@{}c@{}} SVM Classifier Ensemble\cite{vergara2012chemical}\end{tabular}&MLP & \begin{tabular}{@{}c@{}} AddNet-MLP\end{tabular} & \begin{tabular}{@{}c@{}} DiscGAN\end{tabular}  & \begin{tabular}{@{}c@{}} AddNet-DiscGAN \end{tabular} & \begin{tabular}{@{}c@{}} Domain \\adaptation \end{tabular}\\
    \midrule
    Batch 3&87.8& \textbf{98.6} & \textbf{98.6}& 98.3&97.8&98.7\\
    Batch 4&\textbf{90.6}& 83.8 &  75.1&71.4&69.6&73.3\\ 
    Batch 5&72.1& \textbf{99.5} & 99.4&98.4&98.9&99.5\\
    Batch 6&44.5& 74.9 & \textbf{75.9}&72.3&73.9&76.4\\
    Batch 7&42.5& 59.8 & 57.4&61.5&\textbf{66.3}&59.2\\
    Batch 8&29.9& 34.0 & 34.0&\textbf{62.3}&58.8&39.1\\
    Batch 9&59.8& 31.6 & 38.9&63.2&\textbf{63.8}&52.3\\
    Batch 10&39.7& 47.3 & \textbf{54.3}&43.8&44.5&46.1\\
    \bottomrule
\end{tabular}
\label{tab:uci_numerical_table}
\end{table*}

As we can see from Table \ref{tab:uci_numerical_table}, using DiscGAN (with a regular discriminator or AddNet), we were able to obtain better recognition rates for later batches (batches 7, 8 and 9). This could be attributed to the fact that the generator did expose the discriminator to novel unseen points in the data space during training. Therefore, the discriminator would have been able to learn additional meaningful features. As for AddNet, it can perform as well as the regular MLP, either in conventional binary classification or in the case of DiscGAN. It is also worth noting that the domain adaptation scheme we employed did not yield any significant improvements. 
We believe that improved classification results would have been attained, if the entire temporal sensor signal set were at our disposal as input to our algorithms. 
\section{Conclusions}
\label{sec:conclusion}
In this paper, we have introduced a variety of deep-learning based algorithms and applied them to VOC gas and ammonia vapor leak detection and gas type identification problems. The first algorithm is based on AddNet. In AddNet, we replace the computationally expensive dot-product operations in deep neural networks with a modified addition operation that retains the sign of multiplication. Its computational efficiency enables AddNet to be used in embedded and mobile systems, in which we envision a smart gas leakage monitoring and detection CPS being reliably used.

The second algorithm is called DiscGAN, which uses the discriminator of a generative adversarial neural network as a classifier in a bid to enhance the recognition capabilities of the system. The generator part helps in exposing the discriminator to realistic synthetic data points that can be helpful in classification tasks.

We considered three detection and classification tasks. The first task is to detect VOC gas leakage from temporal IR data. Our proposed algorithms achieved accuracy rates of $97-98\%$. The second task we considered is to identify gas types using temporal data of sensor arrays. We were able to attain recognition rates of $96.1-96.5\%$. Our third task was to identify gas types using non-temporal data, where the readings are obtained for the same sensor array over a period of 36 months. The sensor measurements suffer from degradation due to sensor drift.

Although our gas identification accuracy results for the early batches in the last data set were quite high, the degradation incurred in later batches resulted in significant identification accuracy drop. We believe that the non-temporal global features reported for the experiments are highly affected by sensor drift. As a result, the features are not sufficiently expressive of the sensor responses for different gas analyte types. Based on our high recognition rates for two temporal data sets considered in this work, we conclude that using sensor measurements in their temporal presentation, and feeding these recordings into deep neural network algorithms, achieves better performance as these algorithms learn discriminative features by themselves with no need to hand-craft features that could be sensitive to error as in the case of the sensor drift problem. 

\bibliographystyle{IEEEtran}
\bibliography{references}

\end{document}